\documentclass[aps,prl,twocolumn,floatfix,showpacs,reprint,]{revtex4-1}

\usepackage{graphicx}
\usepackage{amsmath, amsfonts, amssymb, bm}

\usepackage{amstext}
\usepackage{mathrsfs}
\usepackage{color}

\begin{document}
\title{Nonlinear Breit-Wheeler Pair Production in a Tightly Focused Laser Beam}
\author{A.~Di Piazza}
\email{dipiazza@mpi-hd.mpg.de}
\affiliation{Max-Planck-Institut f\"ur Kernphysik, Saupfercheckweg 1, D-69117 Heidelberg, Germany}

\begin{abstract}
The only available analytical framework for investigating QED processes in a strong laser field systematically relies on approximating the latter as a plane wave. However, realistic high-intensity laser beams feature much more complex space-time structures than plane waves. Here, we show the feasibility of an analytical framework for investigating strong-field QED processes in laser beams of arbitrary space-time structure by determining the energy spectrum of positrons produced via nonlinear Breit-Wheeler pair production as a function of the background field in the realistic assumption that the energy of the incoming photon is the largest dynamical energy in the problem. A numerical evaluation of the angular resolved positron spectrum shows significant quantitative differences with respect to the analogous result in a plane wave, such that the present results will be also important for the design of upcoming strong laser facilities aiming at measuring this process.
\end{abstract}

\maketitle

The success of QED in vacuum calls for testing the theory thoroughly under more challenging conditions as, e.g., those provided by intense background electromagnetic fields. The typical field scale of QED is set by the so-called critical field of QED: $F_{cr}=m^2/|e|=1.3\times 10^{16}\;\text{V/cm}=4.4\times 10^{13}\;\text{G}$ ($m$ and $e<0$ are the electron mass and charge, respectively, and units with $\hbar=c=4\pi\epsilon_0=1$ and $\alpha=e^2\approx 1/137$ are employed) \cite{Landau_b_4_1982,Fradkin_b_1991,Dittrich_b_1985}. In the presence of an electric field $E\sim F_{cr}$ the vacuum becomes unstable under electron-positron pair production and the interaction energy of the electron magnetic moment with a magnetic field $B\sim F_{cr}$ is comparable with the electron rest energy.

Present high-power optical laser facilities have reached intensities $I_0\sim 10^{22}\;\text{W/cm$^2$}$ \cite{Yanovsky_2008} and upcoming 10-PW facilities aim at $I_0\sim 10^{23}\;\text{W/cm$^2$}$ \cite{APOLLON_10P}. Such technological achievements render lasers a feasible tool for testing QED at field strengths effectively of the order of $F_{cr}$ \cite{Mitter_1975,Ritus_1985,Ehlotzky_2009,Reiss_2009,Di_Piazza_2012,Dunne_2014}. In fact, for a fundamental QED process as nonlinear Breit-Wheeler pair production (NBWPP) by an incoming real photon of four-momentum $k^{\mu}=(\omega,\bm{k})$ and a laser field of amplitude $F_0^{\mu\nu}=(\bm{E}_0,\bm{B}_0)$, the effective field strength in units of $F_{cr}$ at which the process occurs is provided by the quantum nonlinearity parameter $\kappa_0=\sqrt{|(F_{0,\mu\nu}k^{\nu})^2|}/mF_{cr}$ (the metric tensor is $\eta^{\mu\nu}=\text{diag}(+1,-1,-1,-1)$) \cite{Mitter_1975,Ritus_1985,Ehlotzky_2009,Reiss_2009,Di_Piazza_2012,Dunne_2014}. Thus, the strong-field QED regime $\kappa_0\gtrsim 1$ can be entered at a laser intensity of $I_0=10^{23}\;\text{W/cm$^2$}$ if the laser field counterpropagates with respect to a photon of energy $\omega\gtrsim 500\;\text{MeV}$. NBWPP has been thoroughly investigated by approximating the laser field as a plane wave \cite{Reiss_1962,Nikishov_1964,Narozhny_2000,Roshchupkin_2001,Reiss_2009,Heinzl_2010b,Mueller_2011b,Titov_2012,Nousch_2012,Krajewska_2013b,Jansen_2013,Augustin_2014,Meuren_2015,Meuren_2016}. However, the mentioned high intensities can be feasibly attained experimentally only by tightly focusing the laser energy not only in time, which can be accounted for within the plane-wave model, but also in space.

In the present Letter we determine analytically the positron energy spectrum of NBWPP in the presence of a strong laser beam of arbitrary spacetime structure in the realistic assumption that the energy of the incoming photon is the largest dynamical energy in the problem. In fact, in the most efficient laser-photon counterpropagating setup this assumption amounts to require that $\omega\gg m\xi_0\gtrsim m$ \cite{Di_Piazza_2014,Di_Piazza_2015}, where $\xi_0=|e|E_0/m\omega_0$, with $\omega_0$ being the central laser angular frequency. Since even for a Ti:Sa laser ($\omega_0=1.55\;\text{eV}$) of intensity $I_0=10^{23}\;\text{W/cm$^2$}$, it is $\xi_0=150$, the conditions $\omega\gg m\xi_0\gtrsim m$ are automatically fulfilled at $\kappa_0\gtrsim 1$ because for $\omega\approx 500\;\text{MeV}$, it is $\omega/m\approx 10^3$. In \cite{Di_Piazza_2014,Di_Piazza_2015} we have exploited this observation to determine analytically the electron wave function and propagator in a strong laser beam of arbitrary spacetime structure within the Wentzel-Kramers-Brillouin (WKB) approximation by including next-to-leading-order terms of the order of $m\xi_0/\varepsilon$, with $\varepsilon$ being the electron energy. Here, we show that the wave functions found in \cite{Di_Piazza_2014,Di_Piazza_2015} can be feasibly employed for a systematic investigation of strong-field QED processes by obtaining a relatively compact, analytical expression of the positron energy spectrum of NBWPP in a tightly focused laser beam. Moreover, due to the significant quantitative differences between the spectra obtained here numerically in a focused Gaussian beam and those evaluated in a plane wave, the present results will be also useful for the design of upcoming intense laser facilities aiming at measuring NBWPP. It is worth mentioning that effects of the laser spatial focusing in Compton and Thomson scattering have been recently investigated numerically in \cite{Li_2015} and in \cite{Harvey_2016}, respectively. Also, analytical expressions of scalar wave functions based on the WKB approximation have been determined in \cite{Heinzl_2016} for a specific class of background fields depending on the space-time coordinates still via the scalar $(kx)$ like a plane wave but generalizing from lightlike $k^{\mu}$ to arbitrary $k^{\mu}$.

Let us assume that the laser beam is described by the four-vector potential $A^{\mu}(x)$ in the Lorentz gauge $\partial_{\mu}A^{\mu}(x)=0$. If the positive $z$-direction is chosen along the propagation of the incoming photon, it is convenient to employ the light-cone coordinates $T=(t+z)/2$, $\bm{x}_{\perp}=(x,y)$, and $\phi=t-z$ for a generic four-position $x^{\mu}=(t,x,y,z)$. Analogously, we introduce the light-cone components $v_{\pm}=(v^0\pm v_z)/2^{(1\pm 1)/2}$ and $\bm{v}_{\perp}=(v_x,v_y)$ for an arbitrary four-vector $v^{\mu}=(v^0,v_x,v_y,v_z)$. The incoming photon has four-momentum $k^{\mu}=(\omega,0,0,\omega)$ and polarization $l$, whereas the final electron (positron) has four-momentum $p^{\mu}=(\varepsilon,\bm{p})$ ($p^{\prime\mu}=(\varepsilon',\bm{p}')$), with $p^2=m^2$ ($p^{\prime\,2}=m^2$), and spin quantum number $s$ ($s'$). The leading-order $S$-matrix element of NBWPP in the Furry picture reads \cite{Furry_1951,Landau_b_4_1982}
\begin{equation}
S_{fi}=-ie\sqrt{4\pi}\int d^4x\,\bar{\psi}^{(\text{out})}_{p,s}(x)\frac{\hat{e}_{k,l}}{\sqrt{2\omega}}e^{-i(kx)}\psi^{(\text{out})}_{-p',-s'}(x),
\end{equation}
where the hat indicates the contraction of a four-vector with the Dirac matrices $\gamma^{\mu}$, where $e^{\mu}_{k,l}$ is the photon polarization four-vector, and where $\bar{\psi}=\psi^{\dag}\gamma^0$ for an arbitrary bispinor $\psi$. Under the conditions $\omega\gg m\xi_0\gtrsim m$ and $\varepsilon\sim\varepsilon'\sim\omega$, and up to the leading order in $1/\omega$, the external field can be assumed to be independent of $\phi$ and the positive-/negative-energy out-state $\psi^{(\text{out})}_{\pm p,\pm s}(x)$ found in \cite{Di_Piazza_2014,Di_Piazza_2015} can be employed:
\begin{equation}
\label{out_states}
\psi^{(\text{out})}_{\pm p,\pm s}(x)=e^{iS^{(\text{out})}_{\pm p}(x)}\bigg[1\pm\frac{e}{4 p_+}\hat{n}\hat{\mathcal{A}}^{(\text{out})}(\bm{x})\bigg] \frac{u_{\pm p,\pm s}}{\sqrt{2\varepsilon}},
\end{equation}
where
\begin{widetext}
\begin{equation}
S^{(\text{out})}_{\pm p}(x)=\mp(p_+\phi+p_-T-\bm{p}_{\perp}\cdot\bm{x}_{\perp})+e\int_T^{\infty}d\tilde{T} A_-(\tilde{\bm{x}})+\frac{1}{p_+}\int_T^{\infty}d\tilde{T} \left[e(p\mathcal{A}^{(\text{out})}(\tilde{\bm{x}}))\mp
\frac{1}{2}e^2\mathcal{A}^{(\text{out})\,2}(\tilde{\bm{x}})\right],
\end{equation}
\end{widetext}
where $\mathcal{A}^{(\text{out}),\mu}(\bm{x})=(0,\bm{\mathcal{A}}^{(\text{out})}_{\perp}(\bm{x}),0)$, with (see in particular \cite{Di_Piazza_2015})
\begin{equation}
\begin{split}
\bm{\mathcal{A}}^{(\text{out})}_{\perp}(\bm{x})=&\bm{A}_{\perp}(\bm{x})-\bm{\nabla}_{\perp}\int_T^{\infty}d\tilde{T}A_-(\tilde{\bm{x}})\\
=&\int_T^{\infty}d\tilde{T}[\bm{E}_{\perp}(\tilde{\bm{x}})+\bm{z}\times\bm{B}_{\perp}(\tilde{\bm{x}})],
\end{split}
\end{equation}
and $\bm{x}=(T,\bm{x}_{\perp})$ ($\tilde{\bm{x}}=(\tilde{T},\bm{x}_{\perp})$), and where $u_{\pm p,\pm s}$ are the positive-/negative-energy constant free bispinors (a unity quantization volume is assumed) \cite{Landau_b_4_1982}. 

Unlike in the plane-wave case, due to the complex spacetime dependence of the external field on three coordinates, a direct evaluation of the matrix element $S_{fi}$ would not allow for obtaining manageable analytical results. We circumvent this problem by calculating directly the relevant quantity $(1/2)\sum_{l,s,s'}|S_{fi}|^2$. After working out the usual algebra involving traces of the Dirac matrices and by indicating as $\rho_{\Sigma}$ the number of photons impinging into the laser field per unit surface, a lengthy but straightforward calculation provides the number $dN$ of electrons (positrons) created with momenta between $\bm{p}$ ($\bm{p}'$) and $\bm{p}+d\bm{p}$ ($\bm{p}'+d\bm{p}'$) in the form (see the Supplemental Material for a more detailed derivation)
\begin{widetext}
\begin{equation}
\label{dN_1}
\begin{split}
dN=&\rho_{\Sigma}\frac{\pi\alpha}{\omega\varepsilon\varepsilon'}\frac{d^2\bm{p}_{\perp}}{(2\pi)^2}\frac{d\varepsilon'}{2\pi}\frac{d^2\bm{p}'_{\perp}}{(2\pi)^2}\int d^3\bm{x}d^3\bm{x}'e^{i[\Phi(\bm{x})-\Phi(\bm{x}')]}\left\{m^2\left(\frac{\varepsilon'}{\varepsilon}+\frac{\varepsilon}{\varepsilon'}+4\right)+\frac{\varepsilon'}{\varepsilon}\bm{p}^2_{\perp}+\frac{\varepsilon}{\varepsilon'}\bm{p}^{\prime\, 2}_{\perp}-2\bm{p}_{\perp}\cdot\bm{p}'_{\perp}\right.\\
&\left.+e\frac{\omega}{\varepsilon\varepsilon'}(\varepsilon\bm{p}'_{\perp}-\varepsilon'\bm{p}_{\perp})\cdot[\bm{\mathcal{A}}_{\perp}(\bm{x})+\bm{\mathcal{A}}_{\perp}(\bm{x}')]+e^2\left[\bm{\mathcal{A}}^2_{\perp}(\bm{x})+\bm{\mathcal{A}}^2_{\perp}(\bm{x}')+\left(\frac{\varepsilon'}{\varepsilon}+\frac{\varepsilon}{\varepsilon'}\right)\bm{\mathcal{A}}_{\perp}(\bm{x})\cdot\bm{\mathcal{A}}_{\perp}(\bm{x}')\right]\right\},
\end{split}
\end{equation}
where
\begin{equation}
\Phi(\bm{x})=\left(\frac{m^2+\bm{p}^2_{\perp}}{2\varepsilon}+\frac{m^2+\bm{p}^{\prime 2}_{\perp}}{2\varepsilon'}\right)T-(\bm{p}_{\perp}+\bm{p}'_{\perp})\cdot\bm{x}_{\perp}+e\left(\frac{\bm{p}_{\perp}}{\varepsilon}-\frac{\bm{p}'_{\perp}}{\varepsilon'}\right)\cdot\int_T^{\infty}d\tilde{T} \bm{\mathcal{A}}_{\perp}(\tilde{\bm{x}})-\frac{\omega}{\varepsilon\varepsilon'}\frac{e^2}{2}\int_T^{\infty}d\tilde{T} \bm{\mathcal{A}}^2_{\perp}(\tilde{\bm{x}}),
\end{equation}
\end{widetext}
and where we have exploited the conservation law $\omega=\varepsilon+\varepsilon'$ under the approximations $p_+\approx \varepsilon$, $p'_+\approx \varepsilon'$, and $k_+\approx \omega$. Note that the on-shell condition for the electron (positron) implies that $p_-=(m^2+\bm{p}^2_{\perp})/2p_+\approx(m^2+\bm{p}^2_{\perp})/2\varepsilon$ ($p'_-=(m^2+\bm{p}^{\prime 2}_{\perp})/2p'_+\approx(m^2+\bm{p}^{\prime 2}_{\perp})/2\varepsilon'$). Also, we have removed the upper index $(\text{out})$ from the quantity $\bm{\mathcal{A}}^{(\text{out})}_{\perp}(\bm{x})$ for notational simplicity. In order to evaluate the angular resolved positron energy spectrum $dN/d\varepsilon'd\Omega'$, where $d\Omega'\approx d^2\bm{p}'_{\perp}/\varepsilon^{\prime\,2}$, we perform the Gaussian integrals in $d^2\bm{p}_{\perp}$ and the result is (see the Supplemental Material for a more detailed derivation and see also \cite{Baier_b_1998,Dinu_2013})
\begin{widetext}
\begin{equation}
\label{dN_2}
\begin{split}
\frac{dN}{d\varepsilon'd\Omega'}=&i\rho_{\Sigma}\frac{\alpha\varepsilon'}{16\pi^3\omega}\int \frac{d^3\bm{x}d^3\bm{x}'}{T_-}e^{i\Delta\Phi(\bm{x},\bm{x}')}\left\langle m^2\left(\frac{\varepsilon'}{\varepsilon}+\frac{\varepsilon}{\varepsilon'}+4\right)+\frac{2i\varepsilon'}{T_-}+\frac{\varepsilon}{\varepsilon'}\left\{\bm{P}^{\prime}_{\perp}(\bm{x},\bm{x}')-\frac{\varepsilon'}{T_-}\bm{x}_{\perp,-}\right.\right.\\
&\left.\left.+e\frac{\omega}{\varepsilon}\left[\frac{1}{T_-}\left(\int_T^{\infty}d\tilde{T} \bm{\mathcal{A}}_{\perp}(\tilde{\bm{x}})-\int_{T'}^{\infty}d\tilde{T}' \bm{\mathcal{A}}_{\perp}(\tilde{\bm{x}}')\right)+\bm{\mathcal{A}}_{\perp,+}(\bm{x},\bm{x}')\right]\right\}^2-e^2\frac{(\varepsilon-\varepsilon')^2}{4\varepsilon\varepsilon'}\bm{\mathcal{A}}^2_{\perp,-}(\bm{x},\bm{x}')\right\rangle,
\end{split}
\end{equation}
where it turned out to be convenient to introduce the quantities $T_{\pm}=(T\pm T')/2^{(1\pm 1)/2}$, $\bm{x}_{\perp,\pm}=(\bm{x}_{\perp}\pm \bm{x}'_{\perp})/2^{(1\pm 1)/2}$, $\bm{\mathcal{A}}_{\perp,\pm}(\bm{x},\bm{x}')=[\bm{\mathcal{A}}_{\perp}(\bm{x})\pm\bm{\mathcal{A}}_{\perp}(\bm{x}')]/2^{(1\pm 1)/2}$, $\bm{P}'_{\perp}(\bm{x},\bm{x}')=\bm{p}'_{\perp}-eT_-^{-1}\big[\int_T^{\infty}d\tilde{T} \bm{\mathcal{A}}_{\perp}(\tilde{\bm{x}})-\int_{T'}^{\infty}d\tilde{T}' \bm{\mathcal{A}}_{\perp}(\tilde{\bm{x}}')\big]$, and where
\begin{equation}
\label{Delta_Phi}
\begin{split}
\Delta\Phi(\bm{x},\bm{x}')=&\frac{\omega}{\varepsilon\varepsilon'}\frac{T_-}{2}\left[m^2+\bm{P}^{\prime\,2}_{\perp}(\bm{x},\bm{x}')\right]-\frac{\varepsilon}{2T_-}\left[\bm{x}_{\perp,-}+\frac{T_-}{\varepsilon}\bm{P}'_{\perp}(\bm{x},\bm{x}')\right]^2\\
&-\frac{\omega}{\varepsilon\varepsilon'}\frac{e^2}{2}\left\{\frac{1}{T_-}\left[\int_T^{\infty}d\tilde{T} \bm{\mathcal{A}}_{\perp}(\tilde{\bm{x}})-\int_{T'}^{\infty}d\tilde{T}' \bm{\mathcal{A}}_{\perp}(\tilde{\bm{x}}')\right]^2+\int_T^{\infty}d\tilde{T} \bm{\mathcal{A}}^2_{\perp}(\tilde{\bm{x}})-\int_{T'}^{\infty}d\tilde{T}' \bm{\mathcal{A}}^2_{\perp}(\tilde{\bm{x}}')\right\}.
\end{split}
\end{equation}
\end{widetext}
The way how the integral in $d^2\bm{p}_{\perp}$ is performed in Eq. (\ref{dN_1}) indicates how to shift the pole in $T_-$ in Eq. (\ref{dN_2}) and the quantity $T_-$ in the denominator has to be intended as $T_-+i0$ (see the Supplemental Material). Now, we perform the change of variable $d^2\bm{x}_{\perp}d^2\bm{x}'_{\perp}\to d^2\bm{x}_{\perp,+}d^2\bm{x}_{\perp,-}$ and evaluate the integral in $d^2\bm{x}_{\perp,-}$. In fact, from the first two terms in the phase $\Delta\Phi(\bm{x},\bm{x}')$, one sees that for a given positron emission direction, the largest contribution to the integral comes from the region where $|\bm{P}'_{\perp}(\bm{x},\bm{x}')|\lesssim m$. Thus, at energies $\varepsilon\sim\varepsilon'\sim\omega$ where most of the pairs are produced (see also Fig. 1 below), the formation length in $T_-$ is, as in the plane-wave case, of the order of $\varepsilon\varepsilon'/m^2\omega\sim \lambda_0\kappa_0/\xi_0\sim \lambda_C\omega/m$, where $\lambda_0$ is the central laser wavelength and $\lambda_C=1/m=3.9\times 10^{-11}\;\text{cm}$ is the Compton wavelength \cite{Ritus_1985}. This implies that if we set $\bm{X}_{\perp,-}=\bm{x}_{\perp,-}+(T_-/\varepsilon)\bm{P}'_{\perp}(\bm{x},\bm{x}')$, then the formation region in $|\bm{X}_{\perp,-}|$ is of the order of $\lambda_C$ and the same holds for $|\bm{x}_{\perp,-}|$ (see Eq. (\ref{Delta_Phi})). Now, since the field depends either on $\bm{x}_{\perp}=\bm{x}_{\perp,+}+\bm{x}_{\perp,-}/2$ or on $\bm{x}'_{\perp}=\bm{x}_{\perp,+}-\bm{x}_{\perp,-}/2$, we are allowed everywhere to neglect the quantity $\bm{x}_{\perp,-}$ in the field itself because the corrections would be of the order of $\omega_0|\bm{x}_{\perp,-}|\sim\omega_0/m\sim (m/\omega)(\kappa_0/\xi_0)\lesssim m/\omega\ll m\xi_0/\omega\ll 1$. This observation implies that the interference among the contributions to the pair-production probability for different transverse coordinates is already destructive at scales of the order of $\lambda_C$ such that the local constant-field approximation with respect to these coordinates can be safely employed for optical (and x-ray) laser fields. Concerning the coordinate $T_-$, however, the interference among different contributions becomes destructive for $\omega_0|T_-|\sim \kappa_0/\xi_0$, which does not always allow for the use of the local constant-field approximation.

Neglecting the dependence of the external field on $\bm{x}_{\perp,-}$ allows one to perform analytically the resulting Gaussian integral in $\bm{x}_{\perp,-}$ and the relatively compact expression (see the Supplemental Material for a more detailed derivation)
\begin{widetext}
\begin{equation}
\label{Angular}
\begin{split}
\frac{dN}{d\varepsilon'd\Omega'}=&\rho_{\Sigma}\frac{\alpha}{8\pi^2}\frac{m^2}{\varepsilon^2\omega}\int  dT dT' d^2\bm{x}_{\perp}e^{i\frac{\omega}{2\varepsilon\varepsilon'}\left\langle T_-\left\{m^2+\left[\bm{p}'_{\perp}-\frac{e}{T_-}\int_T^{T'}d\tilde{T} \bm{\mathcal{A}}_{\perp}(\tilde{\bm{x}})\right]^2\right\}-e^2\left\{\frac{1}{T_-}\left[\int_T^{T'}d\tilde{T} \bm{\mathcal{A}}_{\perp}(\tilde{\bm{x}})\right]^2+\int_T^{T'}d\tilde{T} \bm{\mathcal{A}}^2_{\perp}(\tilde{\bm{x}})\right\}\right\rangle}\\
&\times\left\{\varepsilon^{\prime\,2}+\varepsilon^2+4\varepsilon\varepsilon'+\frac{\omega^2}{m^2}\left[\bm{p}'_{\perp}+e\frac{\bm{\mathcal{A}}_{\perp}(\bm{x})+\bm{\mathcal{A}}_{\perp}(\bm{x}')}{2}\right]^2-e^2\frac{(\varepsilon-\varepsilon')^2}{m^2}\left[\frac{\bm{\mathcal{A}}_{\perp}(\bm{x})-\bm{\mathcal{A}}_{\perp}(\bm{x}')}{2}\right]^2\right\},
\end{split}
\end{equation}
\end{widetext}
can be obtained, where the index $+$ has been removed from the variable $\bm{x}_{\perp,+}$ for notational simplicity and $\bm{x}'=(T',\bm{x}'_{\perp})\approx (T',\bm{x}_{\perp})$. The positron energy spectrum $dN/d\varepsilon'$ is obtained by integrating with respect to $d\Omega'\approx d^2\bm{p}'_{\perp}/\varepsilon^{\prime\,2}$ and the result is (see the Supplemental Material for a more detailed derivation)
\begin{widetext}
\begin{equation}
\label{Spectral}
\begin{split}
\frac{dN}{d\varepsilon'}=&i\rho_{\Sigma}\frac{\alpha}{4\pi}\frac{m^2}{\varepsilon\varepsilon'\omega^2}\int  \frac{dT dT' d^2\bm{x}_{\perp}}{T_-}e^{i\frac{\omega}{2\varepsilon\varepsilon'}\left\langle m^2T_--e^2\left\{\frac{1}{T_-}\left[\int_T^{T'}d\tilde{T} \bm{\mathcal{A}}_{\perp}(\tilde{\bm{x}})\right]^2+\int_T^{T'}d\tilde{T} \bm{\mathcal{A}}^2_{\perp}(\tilde{\bm{x}})\right\}\right\rangle}\Bigg\{\varepsilon^{\prime\,2}+\varepsilon^2+4\varepsilon\varepsilon'+i\frac{2\omega}{T_-}\frac{\varepsilon\varepsilon'}{m^2}\\
&\left.+e^2\frac{\omega^2}{m^2}\Bigg[\frac{1}{T_-}\int_T^{T'}d\tilde{T} \bm{\mathcal{A}}_{\perp}(\tilde{\bm{x}})+\frac{\bm{\mathcal{A}}_{\perp}(\bm{x})+\bm{\mathcal{A}}_{\perp}(\bm{x}')}{2}\Bigg]^2-e^2\frac{(\varepsilon-\varepsilon')^2}{m^2}\left[\frac{\bm{\mathcal{A}}_{\perp}(\bm{x})-\bm{\mathcal{A}}_{\perp}(\bm{x}')}{2}\right]^2\right\}.
\end{split}
\end{equation}
\end{widetext}
The above results in Eqs. (\ref{Angular}) and (\ref{Spectral}) are valid also for $\xi_0\sim 1$ and are in agreement with the corresponding results obtained in \cite{Baier_b_1998} by means of the operator technique \cite{Baier_1968,Baier_1969,Katkov_2001}. As it should be, the agreement is obtained once the quantities $m\gamma \bm{v}_{0\perp}$ and $m\gamma\Delta\bm{v}(t)$ relative to the electron in Eq. (3.29) in \cite{Baier_b_1998} are identified here with $-\bm{p}'_{\perp}$ and $-e\bm{\mathcal{A}}_{\perp}(\bm{x})$, respectively (note that the term linear in $\Delta\bm{v}(t)$ in Eq. (3.29) in \cite{Baier_b_1998} should have the opposite sign). This casts the operator technique, which
practically allows one to obtain results only at the leading order in the quasiclassical, ultrarelativistic limit and does not contain a general prescription on how to calculate, e.g., high-order corrections, in a general framework where any process can be systematically investigated by employing the Furry picture with the wave functions and the propagator given in \cite{Di_Piazza_2014,Di_Piazza_2015} and by manipulating the resulting analytical expressions as indicated here above. It was also observed in \cite{Akhiezer_b_1996}, where the analogous result for nonlinear Compton scattering in a time-independent external field has been obtained, that the expression of the pair-production probability in \cite{Baier_b_1998} does not contain the average with respect to the coordinates (the transverse coordinates here). More importantly, the results in \cite{Baier_b_1998,Akhiezer_b_1996} are not explicitly expressed in terms of the external field as here but in terms of the electron (positron) trajectories \cite{Baier_b_1998} and  of the classical action \cite{Akhiezer_b_1996}, which have to be determined separately in order to evaluate the angular distribution and the energy spectrum.

As we have hinted above, in the regime $\kappa_0\sim 1$ and $\xi_0\gg 1$ Eqs. (\ref{Angular}) and (\ref{Spectral}) can be further simplified because the external field varies slowly in the formation length $\sim\lambda_0\kappa_0/\xi_0$ of the variable $T_-$ such that the field itself can be expanded around the quantity $T_+$. As a result the integral in $T_-$ can be performed analytically and the results can be expressed in terms of the modified Bessel functions $\text{K}_{\nu}(z)$ \cite{NIST_b_2010} as (see the Supplemental Material for a more detailed derivation)
\begin{equation}
\label{Angular_xi_l}
\begin{split}
\frac{dN}{d\varepsilon'd\Omega'}=&\rho_{\Sigma}\frac{\alpha}{\pi^2\sqrt{3}}\frac{\varepsilon^{\prime\,2}}{\omega^2}\int  d^3\bm{x}\,f(\bm{x})b(\bm{x})\\
&\times\left[1+\frac{\varepsilon^2+\varepsilon^{\prime\,2}}{\varepsilon\varepsilon'}f^2(\bm{x})\right]\text{K}_{1/3}\left(\frac{2}{3}b(\bm{x})f^3(\bm{x})\right)
\end{split}
\end{equation}
and
\begin{equation}
\label{Spectral_xi_l}
\begin{split}
\frac{dN}{d\varepsilon'}=&\rho_{\Sigma}\frac{\alpha}{\pi\sqrt{3}}\frac{m^2}{\omega^2}\int  d^3\bm{x}\Bigg[\frac{\varepsilon^2+\varepsilon^{\prime\,2}}{\varepsilon\varepsilon'}\text{K}_{2/3}\left(\frac{2}{3}b(\bm{x})\right)\\
&\left.+\int_{\frac{2}{3}b(\bm{x})}^{\infty}dz\,\text{K}_{1/3}(z)\right],
\end{split}
\end{equation}
where the index $+$ has been removed from the variable $T_+$ for notational simplicity, where $f(\bm{x})=\sqrt{1+[\bm{p}'_{\perp}+e\bm{\mathcal{A}}_{\perp}(\bm{x})]^2/m^2}$, and where $b(\bm{x})=(\omega^2/\varepsilon\varepsilon')\kappa^{-1}(\bm{x})$, with $\kappa(\bm{x})=(\omega/m)|\partial\bm{\mathcal{A}}_{\perp}(\bm{x})/\partial T|/F_{cr}$ being the local value of the quantum nonlinearity parameter. Both Eq. (\ref{Angular_xi_l}) and Eq. (\ref{Spectral_xi_l}) are in agreement with the corresponding results in \cite{Baier_b_1998} apart from the average on the transverse coordinates but again the advantage here is that also the angular resolved energy spectrum in Eq. (\ref{Angular_xi_l}) is explicitly expressed in terms of the external field. The agreement with the results in \cite{Baier_b_1998} is obtained once, as it should be, the transverse electron momentum there is identified here with $-\bm{p}'_{\perp}-e\bm{\mathcal{A}}_{\perp}(\bm{x})$ and the external field with $-2\partial\bm{\mathcal{A}}_{\perp}(\bm{x})/\partial T$. In this respect, the results in Eq. (\ref{Angular_xi_l}) and Eq. (\ref{Spectral_xi_l}) validate the use of the local version of the constant-crossed-field expressions of the spectra in the regime $\xi_0\gg 1$ not only for a plane wave, as proved in \cite{Ritus_1985}, but also for a spatially focused laser beam under the present approximations. In this respect, we point out that both in the case of $\bm{x}_{\perp,-}$ and of $T_-$ (the latter at $\kappa_0\sim 1$ and $\xi_0\gg 1$) we have taken into account only the (leading) contributions to the integrals coming from the regions $\omega_0|\bm{x}_{\perp,-}|\ll 1$ and $\omega_0|T_-|\ll 1$. If we would have correspondingly evaluated the amplitude $S_{fi}$ by means of the stationary-phase method, this would have been equivalent to take into account in $|S_{fi}|^2$ only the square-modulus of all possible saddle-points and ignoring the highly-oscillating interference terms. As we have already seen in \cite{Meuren_2016} for the plane-wave case, this amounts to ignore the highly-oscillating features of the energy spectra and Eqs. (\ref{Angular_xi_l}) and (\ref{Spectral_xi_l}) actually provide the envelopes of the corresponding highly-oscillating spectra.

In order to point out the importance of spatial focusing effects into the number of produced electron-positron pairs, we compare numerically the results from Eq. (\ref{Angular_xi_l}) with the corresponding quantities in a plane wave. The latter are formally obtained from Eq. (\ref{Angular_xi_l}) by removing the dependence of the field on the transverse coordinates and by setting $\int d^2\bm{x}_{\perp}=\Sigma_{\text{eff}}$, where $\Sigma_{\text{eff}}$ is an appropriate effective transverse surface (see, e.g., \cite{Ritus_1985}). Now, we model the spatially focused laser beam as a linearly polarized, Gaussian beam of spot radius $w_0$, Rayleigh length $z_r=\omega_0w_0^2/2$, and pulse shape $g(T)=\sin^2(\omega_0T/N_L)$ for $\omega_0T\in [0,N_L\pi]$ and $g(T)=0$ elsewhere, with $N_L$ being the number of laser cycles. Following \cite{Salamin_2007}, we work beyond the paraxial approximation, we assume that $N_L\gg 1$ but we keep terms up to order $(w_0/z_r)^4$, reminding that even in the diffraction-limit case ($w_0=\lambda_0$) it is $w_0/z_r=1/\pi\approx 0.32$. Thus, in order to make a fair comparison we can fix $\Sigma_{\text{eff}}$ in such a way that the resulting peak power $\Sigma_{\text{eff}}I_0=\Sigma_{\text{eff}}E_0^2/4\pi$ of the plane wave is the same as that of the Gaussian beam as given in \cite{Salamin_2007}, i.e., $\Sigma_{\text{eff}}=(\pi w_0^2/2)[1+(w_0/2z_r)^2+2(w_0/2z_r)^4]$. In Fig. 1 we compare the quantity $dN/d\varepsilon'd^2\bm{p}'_{\perp}$, where $d^2\bm{p}'_{\perp}\approx \varepsilon^{\prime\,2}d\Omega'\approx\varepsilon^{\prime\,2}\theta'd\theta'd\varphi'$, with $\theta'\approx|\bm{p}'_{\perp}|/\varepsilon'\ll 1$ and $\varphi'=\arctan(p'_y/p'_x)$ (the laser field is polarized along the $x$-direction) in the presence of a Gaussian beam with $w_0=\lambda_0=0.8\;\text{$\mu$m}$ and $N_L=10$ (black continuous curves) with the corresponding plane-wave results (red dashed curves) for different values of $\theta$ and for $\varphi'=0$, i.e., on the laser polarization plane where most of the pairs are produced \cite{Ritus_1985}. The chosen laser peak intensity $5\times 10^{20}\;\text{W/cm$^2$}$ ($\xi_0\approx 10$) and external photon energy $4\;\text{GeV}$ (corresponding to $\kappa_0\approx 0.5$) are within presently available values \cite{Yanovsky_2008,Leemans_2014}, and, for the sake of definiteness, the incoming number of photons $\rho_{\Sigma}$ per unit surface has been set equal to $\Sigma^{-1}_{\text{eff}}$. As it is clearly indicated in Fig. 1, for all different angles the plane-wave results, although include the beam temporal pulse shape, overestimate by at least about an order of magnitude the results in a Gaussian beam in the central part of the spectrum where $\varepsilon\approx\varepsilon'\approx\omega/2$ (notice that typical emission direction angles $\theta'$ are less or of the order of $m\xi_0/\omega\approx 0.07^{\circ}$).
\begin{figure}
\begin{center}
\includegraphics[width=\columnwidth]{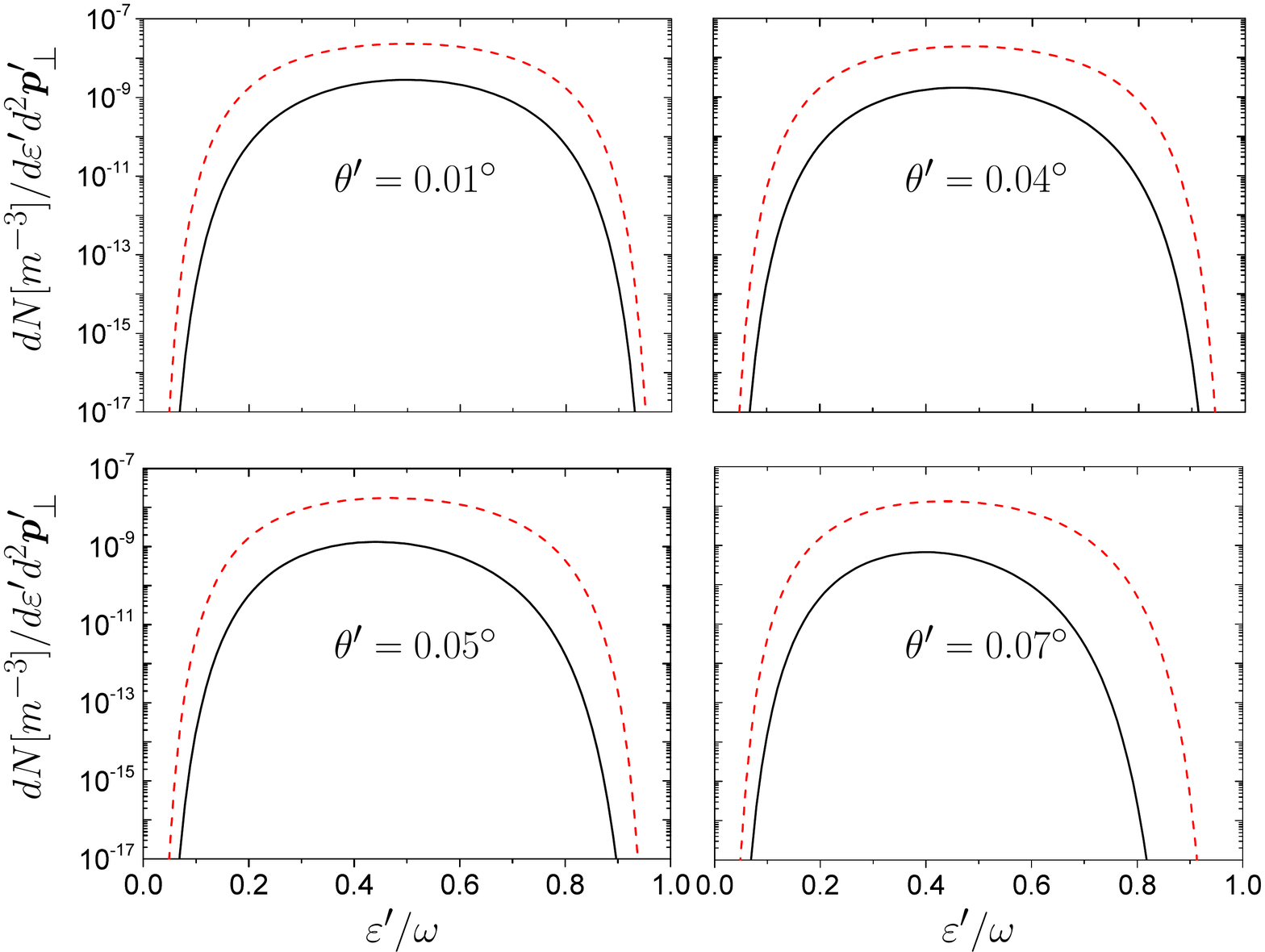}
\caption{Angular resolved positron energy distribution produced via NBWPP in a focused Gaussian beam (black continuous curves) and in a plane wave (red dashed curves) at different values of $\theta'$ and at $\varphi'=0$. See the text for other numerical details.}
\end{center}
\end{figure}

In conclusion, we have put forward a feasible method to systematically investigate strong-field QED processes in virtually arbitrary laser pulses analytically in the experimentally relevant regime where the energy of the incoming particle is the largest dynamical energy in the process. By explicitly investigating NBWPP, we have seen that the tight space focusing of the laser significantly affects the positron spectra and it thus have to be taken into account for the design of upcoming strong laser facilities aiming at scrutinize experimentally strong-field QED.

The author would like to acknowledge useful discussions with A. Angioi, S. Meuren, R. Shaisultanov, and M. Tamburini.

\bibliography{/home/theo/tonywolf/Samba/Travagghiu/Bibliography/Books,/home/theo/tonywolf/Samba/Travagghiu/Bibliography/Reviews,/home/theo/tonywolf/Samba/Travagghiu/Bibliography/Papers_Radiation,/home/theo/tonywolf/Samba/Travagghiu/Bibliography/Papers_RR,/home/theo/tonywolf/Samba/Travagghiu/Bibliography/Papers_PP_and_Cascades,/home/theo/tonywolf/Samba/Travagghiu/Bibliography/Papers_VPE,/home/theo/tonywolf/Samba/Travagghiu/Bibliography/Papers_Crystal,/home/theo/tonywolf/Samba/Travagghiu/Bibliography/Papers_Various}

\end{document}